# Ray-tracing and Interferometry in Schwarzschild Geometry

Farhad Karimi[1], Sina Khorasani[2]

[1]School of Electrical Engineering, Sharif University of Technology, Tehran, Iran
[2]School of Electrical and Computer Engineering, Georgia Institute of Technology
Atlanta, Georgia 30332-0250, USA

**Abstract**— Here, we investigate the possible optical anisotropy of vacuum due to gravitational field. In doing this, we provide sufficient evidence from direct coordinate integration of the null-geodesic equations obtained from the Lagrangian method, as well as ray-tracing equations obtained from the Plebanski's equivalent medium theory. All calculations are done for the Schwarzschild geometry, which results in an anisotropic (pseudo-isotropic) optical equivalent medium when Cartesian coordinates are taken. We confirm that the results of ray-tracing in the equivalent medium and null geodesics are exactly the same, while they are in disagreement with the results of integration in the conventional isotropic equivalent medium of Schwarzschild geometry. Based on the principle invariance of physical due to coordinate transformation, there exist just one result. This Contradiction will be solved by tensor algebra and it will be shown that the conventional isotropic approach is wrong, and even by making isotropic of the metric, the optical behavior of vacuum will remain anisotropic. Hence, we conclude that the true optical behavior of curved spacetime must be anisotropic, and it is an intrinsic property of vacuum in the presence of gravitational field. We provide further discussions on how to detect this possible anisotropy, and what further consequences might be expected in the interpretation of gravtiational lensing data.

## I. INTRODUCTION

Light propagation in gravitational fields [1] has been a matter of intense research during the past decades. Recent experiments of general relativity using the weak gravitational lensing data [2] has revealed that alternative theories of gravity are not excluded. Progress in single atom optics has enabled an accurate tabletop experimental verification of the red shift [3]. Also, advances in the field of artificial metamaterials has established a bridge between the general relativity and optics of anisotropic media [4,5] through the equivalent medium theory [6-8]. This has been newly used to perform microscopic scale tests of propagation under the influence of gravity [9]. An extensive review of the theory and applications of transformation optics has just been published [10].

Recently, a detailed and exact theory of the propagation in the curved space, and in particular, for the Schwarzschild metric has been published by the author [11]. It was shown that the Schwarzschild metric is accompanied by an anisotropy, which inevitably gives rise to an optically anisotropic space with no birefringence, referred to as the pseudo-isotropy. Correction to the Einstein's expression for light deflection was shown and ray-tracing equations were derived. We showed that the time-reversal symmetry of Maxwell's equations breaks down for rotating spacetimes, and through the equivalence principle, we concluded that the light speed should be no longer isotropic for non-inertial frames [11]. Moreover, we had provided some evidence [12], that the anisotropic light propagation in Schwarzschild geometry agrees to Virbhadra's expression [13], by direct calculation of deflection angle in the equivalent medium.

Here, we present simulated images of gravitational lensing by ray-tracing of light beams in both the Equivalent Medium of Isotropic Coordinates (EMIC) and Schwarzschild's Anisotropic Equivalent Medium (SAEM). As it will be shown, the expected anisotropy of space does not affect the appearance of the images, but their apparent sizes. In this way, we establish that the analysis of Einstein's formulation of light deflection actually over-estimates the true mass of blackholes.

Another conclusion drawn from the possible anisotropy of space, as it will be discussed, is the possibility of optical detection of gravity beyond geometrical optics [14]. We also explicitly compare Schwarzschild and gravitational wave metrics, and show that the former is stronger by many orders of magnitude. Hence, we will argue that in principle it would be possible to reveal the existence of a Schwarzschild geometry through a local optical interferometry. For this purpose, the interferometer arms need to rotate by at least 90° in a plane passing through the center of the massive body.

## II. CURVED SPACETIME METRICS

### A. *Exact Calculation of the Index of Refraction*

Here, we only discuss Newtonian and Schwarzschild metrics briefly. The first is given by

$$ds^2 = -c^2\left(1 - 2\frac{r_s}{r}\right)dt^2 + \left(dx^2 + dy^2 + dz^2\right) \quad (1)$$

where $r_s = 2GM/c^2$ is the Schwarzschild radius of the star with $M$ and $G$ respectively being its mass and gravitational constant, and $dl^2 = dx^2 + dy^2 + dz^2$ is the spacelike path element. Denoting the gravitational potential by $\Phi = -r_s/r$, the resulting refractive index is [11]

$$n = (1 + 2\Phi)^{-\frac{1}{2}} \quad (2)$$

In the limit of small $|\Phi| \approx 0$ or $r \gg r_s$, it becomes [15]

$$n \approx 1 - \Phi \quad (3)$$

This is the Einstein's 1911 early result. In contrast, the Schwarzschild metric is [16, p. 607]

$$ds^2 = -c^2\left(1 - \frac{r_s}{r}\right)dt^2 + \quad (4)$$

$$\frac{dr^2}{1 - \frac{r_s}{r}} + r^2\left(d\theta^2 + \sin^2\theta d\phi^2\right)$$

where $(r, \theta, \phi)$ are the standard spherical coordinates. Unlike (1), the metric (4) is evidently anisotropic, in which the spacelike path element $dl^2 = dr^2 + r^2\left(d\theta^2 + \sin^2\theta d\phi^2\right)$ does not appear explicitly. The common practice is to transform (5) using the so-called isotropic coordinates [11,16] with the non-physical radial coordinate $\rho$ defined by $r = \rho\left(1 + \frac{r_s}{4\rho}\right)^2$, which has the inverse relation $\rho = \frac{r_s}{2}\left[\frac{r}{r_s} - \frac{1}{2} + \sqrt{\left(\frac{r}{r_s}\right)^2 - \frac{r}{r_s}}\right]$ for the exterior of the blackhole, to obtain the well-known result for the new metric and refractive index

$$ds^2 = -c^2\left(\frac{1 - \frac{r_s}{4\rho}}{1 + \frac{r_s}{4\rho}}\right)^2 dt^2 + \quad (5a)$$

$$\left(1 + \frac{r_s}{4\rho}\right)^4\left[d\rho^2 + \rho^2\left(d\theta^2 + \sin^2\theta d\phi^2\right)\right]$$

$$n = \left(1 - \frac{r_s}{4\rho}\right)^{-1} \left(1 + \frac{r_s}{4\rho}\right)^{3} \quad (5b)$$

$$= \left\{1 - \frac{1}{2}\left[\frac{r}{r_s} - \frac{1}{2} + \sqrt{\left(\frac{r}{r_s}\right)^2 - \frac{r}{r_s}}\right]^{-1}\right\}^{-1}$$

$$\left\{1 + \frac{1}{2}\left[\frac{r}{r_s} - \frac{1}{2} + \sqrt{\left(\frac{r}{r_s}\right)^2 - \frac{r}{r_s}}\right]^{-1}\right\}^{3}$$

Comparing to (3) reveals a significantly more complicated relation for the refractive index [11,16]. But as we have already discussed in detail [11], isotropic coordinates eventually disregards the physical anisotropy of (4). This normally could lead to physical implications, which are at best partially correct.

In order to illustrate this, we have obtained the *exact* refractive index directly in Schwarzschild coordinates [11]. The refractive index in the equivalent medium of non-rotating vacuum curved spacetime can be calculated by [11]:

$$n = \sqrt{\hat{k} \cdot [\xi] \cdot \hat{k}}^{-1} \quad (6)$$

where $\hat{k}$ is the unit vector along the propagation vector, $[\xi] = [\varepsilon]/|\varepsilon| = [\mu]/|\mu|$, in which $[\varepsilon]$ and $[\mu]$ are respectively permittivity and permeability tensors given by

$$[\varepsilon] = [\mu] = -\frac{\sqrt{-g}}{g_{00}}\left[g^{ij}\right] \quad (7)$$

Here, $g^{ij}$ and $g_{ij}$ are respectively the contravariant and covariant elements of the metric tensor of space, with $g$ being the determinant of 4-metric. For finding the exact refractive index in the Schwarzschild metric, first we rewrite the corresponding metric (4) as

$$ds^2 = -\left(1 - \frac{r_s}{r}\right)dt^2 + \left(\frac{r_s}{r-r_s}\frac{x_i x_j}{r^2} + \delta_{ij}\right)dx^i dx^j \quad (8)$$

At each point, the coordinates can be chosen in a way that the center of the blackhole is at the origin and $\hat{r} = \hat{z}$. Then we have:

$$[g_{\mu\nu}] = \begin{pmatrix} -\left(1 - \frac{r_s}{r}\right) & 0 & 0 & 0 \\ 0 & 1 & 0 & 0 \\ 0 & 0 & 1 & 0 \\ 0 & 0 & 0 & \frac{1}{1-\frac{r_s}{r}} \end{pmatrix} \quad (9)$$

It can be easily seen that $g = -1$. The metric tensor with contravariant elements is also given by

$$[g^{\mu\nu}] = [g_{\mu\nu}]^{-1} = \begin{pmatrix} \frac{-1}{1-\frac{r_s}{r}} & 0 & 0 & 0 \\ 0 & 1 & 0 & 0 \\ 0 & 0 & 1 & 0 \\ 0 & 0 & 0 & 1-\frac{r_s}{r} \end{pmatrix} \quad (10)$$

From (7), permittivity and permeability tensors can be calculated, resulting in

$$[\varepsilon] = [\mu] = \frac{1}{1-\frac{r_s}{r}}\begin{pmatrix} 1 & 0 & 0 \\ 0 & 1 & 0 \\ 0 & 0 & 1-\frac{r_s}{r} \end{pmatrix} \quad (11)$$

Now $[\xi]$ is found to be

$$[\xi] = \frac{[\varepsilon]}{|\varepsilon|} = \frac{[\mu]}{|\mu|} = \begin{pmatrix} 1-\frac{r_s}{r} & 0 & 0 \\ 0 & 1-\frac{r_s}{r} & 0 \\ 0 & 0 & \left(1-\frac{r_s}{r}\right)^2 \end{pmatrix} \quad (12)$$

By inserting (12) in (6), the refractive index resulting from the equivalent medium of vacuum in Schwarzschild metric takes the form

$$|n(\mathbf{r})| = (1+\Phi)^{-\frac{1}{2}}\left(1+\Phi\cos^2\psi\right)^{-\frac{1}{2}} \quad (13)$$

where the angle $\psi$ is made by the radius $\mathbf{r}$ and wavevector $\mathbf{k}$. This expression is evidently direction-dependent, which is due to the anisotropy of the equivalent medium, and also very different to the relation (5b). It should be pointed out, however, that the equivalent medium of space-time is always a particular case of the general anisotropic media, referred to as pseudo-isotropic with vanishing birefringence [11]. For a pseudo-isotropic medium the ordinary and extraordinary rays are always the same and therefore degenerate along all possible directions, and there are no such optical axes at all. This issue has been discussed deeply in [11].

Since light-rays travel along the null geodesics with $ds^2 = 0$ and the phase velocity $v_p^2 = dl^2/dt^2$, the corresponding refractive index would be simply given by $n^2 = c^2/v_p^2 = c^2 \left(dl^2/dt^2\right)^{-1}$. This enables us to readily check the consistency of (13) by inspection of the metric (4) for the two orthogonal directions $\hat{k} \parallel \hat{r}$ and $\hat{k} \perp \hat{r}$. For this purpose we exploit the spherical symmetry in such a way that $\hat{r} = \hat{z}$. Then along the parallel direction $\hat{k} \parallel \hat{r}$ we have $\theta = 0$ and $d\theta = 0$, and therefore $dl^2 = dr^2$. This is while along the perpendicular direction $\hat{k} \perp \hat{r}$ we have $\theta = \frac{\pi}{2}$ and $dr = 0$, and therefore $dl^2 = r^2 \left(d\theta^2 + \sin^2\theta d\phi^2\right)$. Now by comparing to the metric (4) and under these two conditions of parallel and perpendicular propagation, we respectively obtain the refractive indices

$$n_\parallel = \left(1 - \frac{r_s}{r}\right)^{-1}, \quad n_\perp = \left(1 - \frac{r_s}{r}\right)^{-\frac{1}{2}} \quad (14)$$

with $n_\parallel$ and $n_\perp$ coinciding with the conditions $\psi = 0$ and $\psi = \frac{\pi}{2}$ in (13). It can be evidently seen that the vacuum behaves anisotropic, which is certainly not a coordinate artifact.

Now, for small $\Phi$ from (14) we get

$$n \approx 1 - \frac{1 + \cos^2\psi}{2}\Phi \quad (15)$$

to be compared with the Einstein's famous result $n \approx 1 - \Phi$. The correction factor of two as compared with (3), hence actually varies between 1 and 2 depending on the propagation angle.

This anisotropy is present everywhere around a massive object, so that the maximum change in $n$ by changing $\psi$ could reach as high as $|\Phi|/2 = \left(GM/c^2\right)/r$. Based on the available estimates [16, p. 459] and for an experiment at Earth's distance from Sun, this figure is of the order of $10^{-8}$, while it would be only about $6 \times 10^{-10}$ at the surface of Earth when the gravity of Sun is neglected. We later argue in Sec. IV that a local interferometry similar to the Laser Interferometer Gravitational wave Observatory (LIGO) could possibly reveal the existence of Schwarzschild metric.

**B.** *Physical Justifications*

Now let us consider the reason behind this remarkable difference between (5b) and (13). Based on the equivalence principle, one may view the gravitational field as an acceleration field. The relation (14) simply states that the isotropy in the velocity of light does not hold for an inertial observer. This fact may be regarded as an extension of the second postulate of special relativity [11]. The transformation using isotropic coordinates has somehow neglected this extension, since it virtually removes this acceleration caused by gravity, and makes the space locally isotropic. It means that in the isotropic equivalent medium of isotropic coordinates, the direction of acceleration could not be distinguished, sensed, or felt, locally; any measurement of the acceleration of gravity would need a non-local experiment which measures the gradients. This is clearly a not satisfactory explanation, as we know that acceleration is sensed point-wise and therefore locally. Also it will be showed that the refractive index in (5b) will results in wrong prediction of behavior of light in the presence of gravitational field.

A possible argument is that transformations are only mathematical tools to make problems at hand easier to solve. Back transformations should therefore return always to the original answers, just in the same way we employ pure rotations of axes, or transformations using spherical polar, or cylindrical systems of coordinates instead of Cartesian coordinates to solve multi-dimensional electrostatic problems.

This has not happened as one may easily distinguish the difference between (5b) and (14). It might be here argued that an inertial observer must be equivalent to a non-inertial observer since any differential motion of a non-inertial observer along its worldline trajectory could always be placed over the tangent line at that point, which is in inertial motion, and thereby, inertial and non-inertial observers could be point-by-point equivalent. We believe that such an interpretation is also wrong, since the non-inertial observer does not feel the presence of an inward acceleration.

Another equivalent interpretation to the above statement is that, based on the equivalence principle, Minkowskian spacetime can be equivalent to any curved spacetime, through a proper *local* transformation. This does not clearly imply that the spacetime at the observation point is void of any acceleration: bending of light rays and motions of massive objects in central gravitational fields are vivid counter-examples. The acceleration could be sensed in a reference frame only if physically measurable coordinates measure the dimensions around the non-inertial observer.

Another important issue to be pointed out here, is that the non-isotropy of light velocity in a gravitational field as predicted by (14) is a pure local effect, which bears no non-local interpretation. To be explicit, *non-isotropy of light velocity in an accelerated frame is still a local effect*. While this also justifies the hypothesis of locality in the theory of special relativity [19-21], it apparently violates that particular interpretation of equivalence principle, which states that "at each instant along its world line an accelerated observer is [always] equivalent to an otherwise identical momentarily comoving inertial observer" [19-21].

It is worth here mentioning that non-local theories of general relativity [22,23] are based on a gravitational potential which is non-locally and non-linearly defined through an integration over the whole space. It is predicted that these non-local contributions may be able to eventually explain the dark matter.

This discussion puts us to a new debate in this realm; the direct formulation in Schwarzschild coordinates results in the conclusions that:

(a) the gravitational spacetime is optically anisotropic;
(b) the acceleration causes anisotropy in the light velocity;
(c) the interpretation of blackhole mass based on the gravitational lensing data needs corrections due to the anisotropy of space. This is due to the over-estimated angle of deflection caused by the non-physical isotropic space, and will be discussed in the next section;
(d) the possible anisotropy in the light velocity could be revealed by interferometry. However, if any such interferometry fails, then the LIGO experiment would be eventually unable to detect gravitational waves; this point will be discussed in Sec. IV.

# **III. RAY-TRACING USING TRANSFORMATION OPTICS**

### A. *Theory of Ray-tracing*

In equivalent medium theory, the exact equations of motion take the form [4,22]:

$$\frac{d\mathbf{r}}{dl} = +\frac{\partial H}{\partial \mathbf{k}} \quad (16)$$
$$\frac{d\mathbf{k}}{dl} = -\frac{\partial H}{\partial \mathbf{r}}$$

Here, the Hamiltonian $H$ is expressed by $H = f(\mathbf{r})(\mathbf{k} \cdot [\varepsilon] \cdot \mathbf{k} - |\varepsilon|)$. $f(\mathbf{r})$ is an arbitrary function $\mathbf{r}$ of which we choose it $|\varepsilon|^{-1}$ for simplicity. So the Hamiltonian will be $H = \mathbf{k} \cdot \left( |\varepsilon|^{-1} [\varepsilon] \right) \cdot \mathbf{k} - 1$. Also, $dl$ is the differential of an arbitrary quantity proportional to the differential of length along the propagation. Any possible physical meaning of $dl$ therefore depends on the choice and dimension of $f(\mathbf{r})$, but will be nevertheless irrelevant to the final trajectory and deflection angle, which we need to calculate.

Now, we can furthermore notice that due to (6) and with the particular choice of $f(\mathbf{r}) = |\varepsilon|^{-1}$, the Hamiltonian can be expressed concisely by

$$H = k^2 n^{-2} - 1 \tag{17}$$

in which $n$ is the refractive index of equivalent medium.

In Schwarzschild geometry, ray-tracing equations can be obtained as:

$$\frac{d}{dl}\mathbf{S} = \begin{bmatrix} y_{11} & y_{12} \\ y_{21} & y_{22} \end{bmatrix} \mathbf{S} \tag{18}$$

$$y_{11} = 2\frac{k\Phi}{r}(1+\Phi)\cos\psi$$
$$y_{12} = 2(1+\Phi)$$
$$y_{21} = \frac{k^2\Phi}{r^2}\left[1 + (3+4\Phi)\cos^2\psi\right]$$
$$y_{22} = -\frac{2k\Phi}{r}(1+\Phi)\cos\psi$$

in which $\mathbf{S} = \{\mathbf{r} \quad \mathbf{k}\}^T$.

Now, for small $\Phi$ and from the above we get:

$$\frac{d}{dl}\mathbf{S} \approx \begin{bmatrix} 2\frac{k\Phi}{r}\cos\psi & 2(1+\Phi) \\ \frac{k^2\Phi}{r^2}(1+3\cos^2\psi) & -\frac{2k\Phi}{r}\cos\psi \end{bmatrix}\mathbf{S} \tag{19}$$

$$= \begin{bmatrix} 0 & 2 \\ 0 & 0 \end{bmatrix}\mathbf{S} + \Phi\begin{bmatrix} \frac{2k}{r}\cos\psi & 2 \\ \frac{k^2}{r^2}(1+3\cos^2\psi) & -\frac{2k}{r}\cos\psi \end{bmatrix}\mathbf{S}$$

Through an analysis similar to what is done in [11], it is fairly easy to show that the ray-tracing equations for the Isotropic equivalent medium (5b) simply are:

$$\frac{d}{dl}\mathbf{S} = \begin{bmatrix} 0 & 2\frac{\left(1-\frac{r_s}{4\rho}\right)^2}{\left(1+\frac{r_s}{4\rho}\right)^6} \\ \frac{2k^2\Phi}{r^2}\frac{1-\frac{r_s}{8\rho}}{\left(1+\frac{r_s}{4\rho}\right)^4} & 0 \end{bmatrix}\mathbf{S} \tag{20}$$

Now, for small $\Phi$, from (19) we get:

$$\frac{d}{dl}\mathbf{S} \approx \begin{bmatrix} 0 & 2(1+2\Phi) \\ \frac{2k^2\Phi}{r^2} & 0 \end{bmatrix}\mathbf{S} \tag{21}$$

$$= \begin{bmatrix} 0 & 2 \\ 0 & 0 \end{bmatrix}\mathbf{S} + \Phi\begin{bmatrix} 0 & 4\Phi \\ \frac{2k^2\Phi}{r^2} & 0 \end{bmatrix}\mathbf{S}$$

It should be added that both (19) and (21) are correct to $O(\Phi)$. In the absence of gravitational field with $\Phi = 0$, both sets of ray-tracing equations relax to that of a straight light beam moving along $\hat{k}$.

Hereby, we will perform integrations of both sets of ray-tracing equations and compare the results. Equations (19) and (21) may be easily integrated in 2D using the initial conditions $\mathbf{r}(0) = -50\hat{x} - d\hat{y}$ and $\mathbf{k}(0) = k\hat{x}$, since the ray moves on the plane $(\mathbf{r}(0), \mathbf{k}(0))$. The initial position vector is supposed to be enough distant in where both isotropic and anisotropic Schwarzschild metrics relax to the Minkowskian form. Clearly, the closest distance of approach is $d = r_0$.

**B. Results**

Numerical integration of (19) and (21) needs an accurate integration scheme. We actually noted that simple Euler's integration scheme is far more than insufficient for this purpose. Hence, a fourth-order Runge-Kutta method was implemented to achieve stable and reliable results. Calculations were done using a code written in C++ language for maximal performance.

Light rays passing nearby a black hole are illustrated in Fig. 1. Since according to (19) and (21), light rays always lies on a single plane passing through blackhole's center, ray-tracing on a 2D plane is sufficient to notice the deflection of light rays. Fig. 1a shows the light trajectories due to the equivalent medium of isotropic coordinates, while Fig. 1b shows the same for the SAEM. Calculated angles of deflection were noticed to be in rough agreement

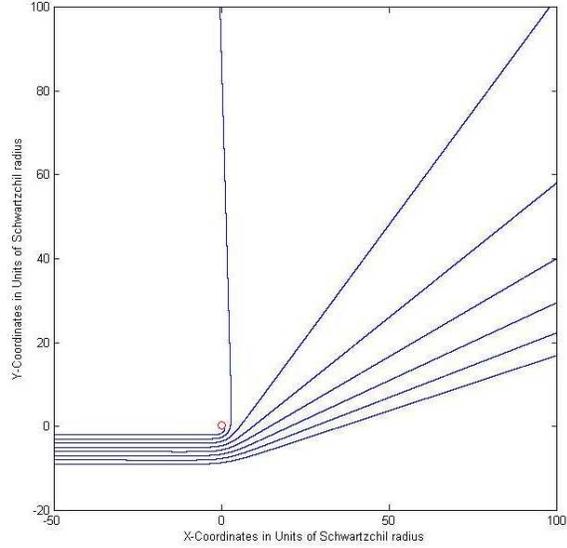

(a)

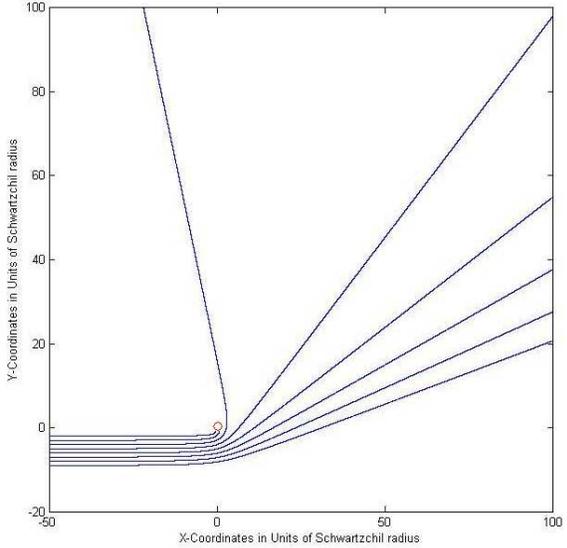

(b)

**Figure 1**. Ray-tracing in space around a blackhole with radius $r_s$ (shown as the red circle): (a) SAEM; (b) EMIC. Spatial dimensions are normalized to $r_s$.

with Einstein's formula for the angle of light deflection [16], given by $\alpha \approx r_s/r_0$.

It could be seen that EMIC causes stronger deflection of light rays. This may be easily understood by referring to (15), which shows that the true refractive index varies between $1 + \frac{1}{2}|\Phi|$ and $1 + |\Phi|$ in the Schwarzschild space, while it is always equal to $1 + |\Phi|$ for the equivalent medium of isotropic coordinates. In other words, the light beam feels a smaller average refractive index in the Schwarzschild space, and therefore undergoes less deflection.

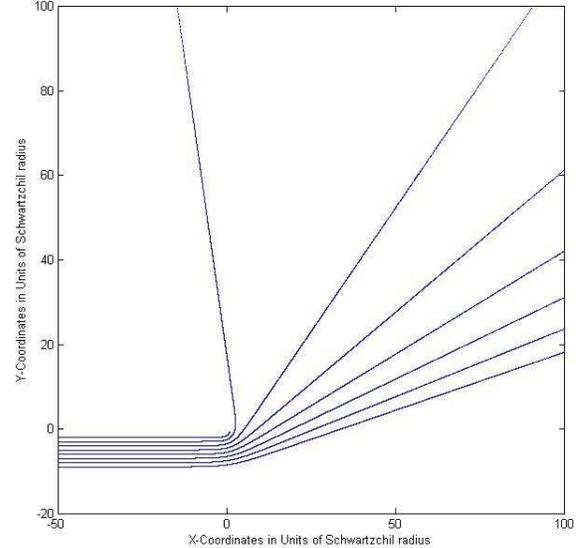

(a)

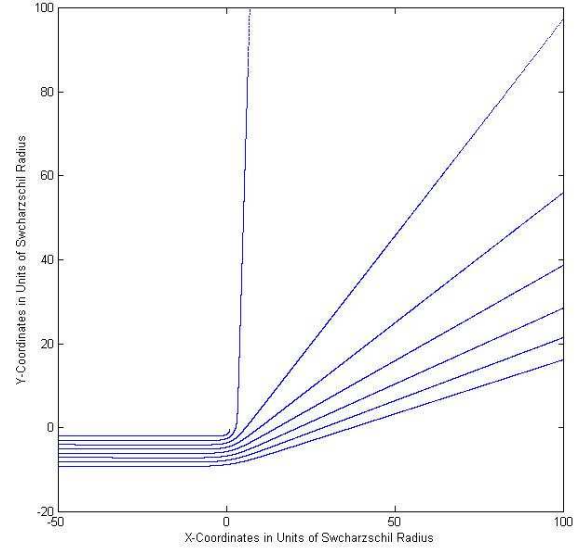

(b)

**Figure 2**. Null-geodesics around a blackhole with radius $r_s$: (a) Anisotropic Schwarzschild Metric; (b) Isotropic Schwarzschild Metric. Spatial dimensions are normalized to $r_s$.

Here we also want to compare the results in equivalent media with null-geodesics in the Schwarzschild and the isotropic metrics. To obtain the geodesic equation we use the

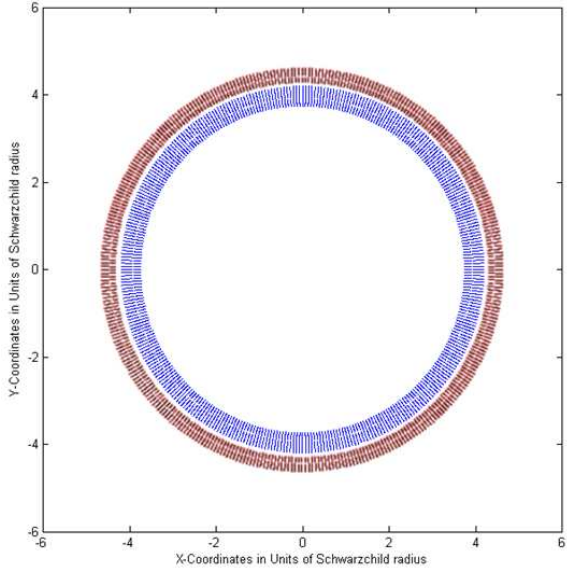

**Figure 3**. Einstein ring of an on-axis star with unit radius: (red) EMIC; (blue) SAEM. In both cases, the blackhole, star, and observer are located respectively at (0,0,0), (0,0,–5), and (0,0,+5). Spatial dimensions are normalized to $r_s$.

Lagrangian method [23]. In the Schwarzschild metric, Lagrangian is given by $L = g_{\mu\nu}\dot{x}^\mu \dot{x}^\nu$, where $\dot{x}^\mu \equiv dx^\mu/d\sigma$. Hence, we easily have

$$L = -\left(1-\frac{r_s}{r}\right)\dot{t}^2 + \frac{\dot{r}^2}{\left(1-\frac{r_s}{r}\right)} + r^2\left(\dot{\theta}^2 + \dot{\varphi}^2 \sin^2\theta\right) \quad (22)$$

For a photon moving on null-geodesics we have $L=0$, and also due to spherical symmetry we can assume $\theta = \pi/2$. Therefore, we get

$$L = -\left(1-\frac{r_s}{r}\right)\dot{t}^2 + \frac{\dot{r}^2}{\left(1-\frac{r_s}{r}\right)} + r^2\dot{\varphi}^2 = 0 \quad (23)$$

The geodesic equations is found by substituting $L$ into the Euler-Lagrange equations given as $d/d\sigma\left(\partial L/\partial \dot{x}^\mu\right) = \partial L/\partial x^\mu$, resulting in

$$\frac{d}{d\sigma}\left[-2\left(1-\frac{r_s}{r}\right)\dot{t}\right] = 0, \quad \mu = 0 \quad (24)$$

$$\frac{d}{d\sigma}\left[2r^2\dot{\varphi}\right] = 0, \quad \mu = 3 \quad (25)$$

The first equation gives $\left(1-\frac{r_s}{r}\right)\dot{t} = k$, while the second results in $r^2\dot{\varphi} = h$, where $k$ and $h$ are constants. By substituting (24) and (25) in (23), we have

$$-\left(1-\frac{r_s}{r}\right)^{-1}k^2 + \left(1-\frac{r_s}{r}\right)^{-1}\dot{r}^2 + \frac{h^2}{r^2} = 0 \quad (26)$$

By substituting $\dot{r} = dr/d\sigma$ with $\dot{\varphi}\, dr/d\varphi = \left(h/r^2\right) dr/d\varphi$ and a bit of simplifying, the geodesic equation is found to be

$$\frac{1}{r^4}\left(\frac{dr}{d\varphi}\right)^2 + \frac{1}{r^2}\left(1-\frac{r_s}{r}\right) = \frac{k^2}{h^2} \quad (27)$$

We now differentiate this equation with respect to $\varphi$, to obtain finally [3]

$$\frac{d^2r}{d\varphi^2} = r - \frac{3}{2}r_s + \frac{2}{r}\left(\frac{dr}{d\varphi}\right)^2 \quad (28)$$

The null-geodesic equation for the isotropic metric can be derived similarly

$$\frac{d^2\rho}{d\varphi^2} = -\frac{\left[\left(\frac{d\rho}{d\varphi}\right)^2 + \rho^2\right]\left(\frac{r_s}{\rho} - 2\right)}{\rho\left[1-\left(\frac{r_s}{4\rho}\right)^2\right]} - \rho \quad (29)$$

By using fourth-order Runge-kutta method, (27) and (28) are integrated and the null-geodesics in the Schwarzschild metric (Fig. 2a) and isotropic metric are obtained (Fig. 2b).

Here by comparing the Fig. 1 and Fig. 2, we see that trajectories of light in Schwarzschild anisotropic equivalent medium, null-geodesics in Schwarzschild metric and also the isotropic metric follow the same trajectories within slight difference, which may be attributed to the error of numerical integration. But the trajectories of light in EMIC are totally different. Here a question arises that why the trajectories of light should be different in spite of invariance of optical behavior of space-time due to coordinate transformation? Here follows an explanation.

Since trajectories of light should be invariant under coordinate transformation, it can be deduced that the Hamiltonian in (16) should be invariant too. But it can be easily proved that in 3+1−dimensions if the coordinate transformation is time invariant then the Hamiltonian is invariant under the coordinate transformation. It means that the Hamiltonian is scalar. Based on (16) with $f(\mathbf{r}) = |\varepsilon|^{-1}$, here we rewrite the Hamiltonian in equivalent medium:

$$H = \mathbf{k} \cdot \left( |\varepsilon|^{-1} [\varepsilon] \right) \cdot \mathbf{k} - 1 \qquad (30)$$

Based on the Plebanski's constitutive relation (7), the tensorial representation of Hamiltonian of light in vacuum and curved space-time will be:

$$H = g_{00} k_i g^{ij} k_j \qquad (31)$$

By applying isotropic coordinate transformation $x^\mu \to x'^\mu$ the Hamiltonian $H'$ will be equal to $H$ due to the scalar nature of the Hamiltonian. Therefore, the Hamiltonian of light propagation in the EMIC is simply given by:

$$H' = g_{00}' k_i' g'^{ij} k_j' = \mathbf{k}' \cdot \left( |\varepsilon'|^{-1} [\varepsilon'] \right) \cdot \mathbf{k}' \qquad (32)$$

Here, we may point out the reason of difference. It is a common mistake that by transforming coordinates, we just change the permittivity matrix in the Hamiltonian and the wavevectors remain unchanged, but here we can see that for the case of isotropic coordinate transformation and in polar coordinates we should take

$$k_i' = \frac{\partial x^i}{\partial x'^i} k_i \qquad (33)$$

which means that the correct derivative is instead

$$\mathbf{k}' = \left( \frac{\partial r}{\partial \rho} k_r \quad k_\theta \quad k_\varphi \right) \qquad (34)$$

Now by substituting the new wavevector in (6) for the isotropic Schwarzschild metric, the refractive index of vacuum can be derived

$$n = \frac{1}{\sqrt{(1+\Phi)(1+\Phi \cos^2 \psi)}} \qquad (13)$$

recovering exactly the same refractive index of vacuum for the Schwarzschild metric as previously had been found in (13) through direct method. It means that even by recasting the Schwarzschild metric into the isotropic form, the *exact* refractive index of vacuum will still remain anisotropic.

By correcting the wavevector in the EMIC, the expected trajectories of light will be corrected too. Then, we may observe that the trajectories of light will be the same with the trajectories of light in anisotropic Schwarzschild equivalent medium. And therefore, there is no contradiction between results and principle of invariance of physical results due to coordinate transformation. By the way, it is obvious that whether the equivalent medium is optically anisotropic or isotropic, the Hamiltonian is dependent to the direction of light. As a result, we can deduce with confidence that optical behavior of vacuum around a spherically symmetric, static and non-rotating massive body is anisotropic.

Here we investigate some consequences of anisotropic optical behavior vacuum around a blackhole. For this reason, so we have simulated the images of gravitational lensing caused by a blackhole located at the origin. We consider three cases: (i) the star is located on-axis and lying at the observation axis (which is the line passing through the observer and blackhole center), (ii) the star is located slightly off-axis yet having an overlap with the observation axis, and (iii) the star is located fully off-axis having no overlap with the observation axis. In all simulations, the spatial coordinates were normalized to the blackhole radius $r_s$, such that the blackhole's radius becomes unity. For simplicity, the star's radius is also taken as unity and is located within 5 unit distances of the blackhole center. The observation point is also at the distance of 5 units away from the blackhole center, on the opposite side.

(i) Fig. 3: Blackhole is located at (0,0,0), star is located at (0,0,-5), and the observation point is at (0,0,+5). Because of the rotational symmetry with respect to the z-axis, the image would be also centrosymmetric. This will be seen in the form of the so-called Einstein's ring. The ring forms in both isotropic (red) and anisotropic (blue) spaces, but the image caused by the EMIC is about 10% larger than the actual image obtained in the SAEM.

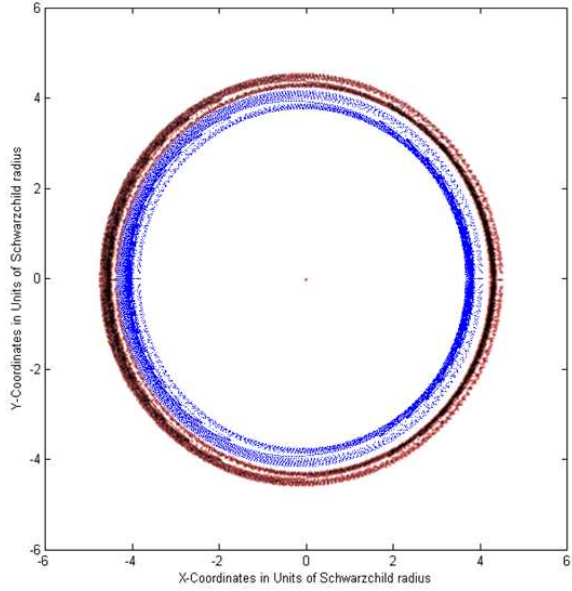

**Figure 4**. Lensing image of an off-axis star with unit radius: (red) EMIC; (blue) SAEM. In both cases, the blackhole, star, and observer are located respectively at (0,0,0), (0,0,–5), and (0,0,+5). Dimensions are normalized to $r_s$.

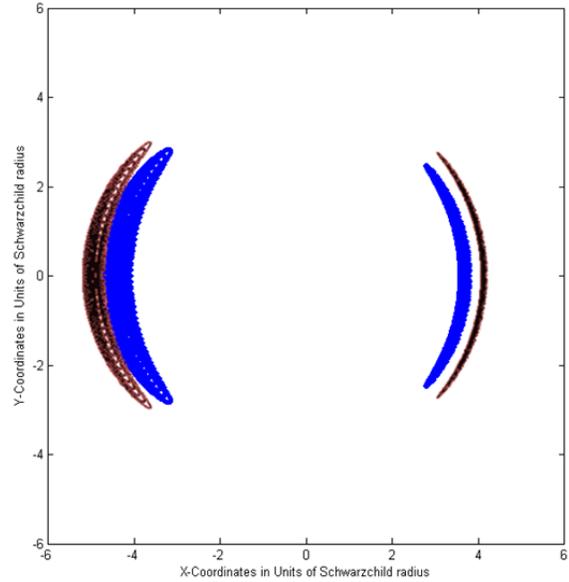

**Figure 5**. Lensing image of an off-axis star with unit radius: (red) EMIC; (blue) SAEM. In both cases, the blackhole, star, and observer are located respectively at (0,0,0), (0,0,–5), and (0,0,+5). Dimensions are normalized to $r_s$.

(ii) Fig. 4: Blackhole is located at (0,0,0), the star at (+0.5,0,-5), and the observation point at (0,0,+5). Since the star has a unit radius, it has intersection with the observation axis, which is the same as z-axis. The configuration has no rotational symmetry, and hence the image is not centrosymmetric. Still the ring is unbroken, but the intensity of light on the ring is not uniform. Furthermore, the center of image has moved slightly to the right. Again, the images of isotropic (red) and anisotropic (blue) spaces look the same in appearance, but the image of the EMIC is about 10% larger than that obtained in the SAEM.

(iii) Fig. 5: Blackhole is located at (0,0,0), the star at (+1.5,0,–5), and the observation point at (0,0,+5). Since the star has a unit radius, it has no intersection with the observation axis, i.e. z-axis. Again, the configuration has no rotational symmetry, and hence the image is not centrosymmetric. But the Einstein's ring is broken into two unequal crescents. Also, the center of image has moved slightly to the right. Again, the images of isotropic (red) and anisotropic (blue) spaces look very similar, but the image of the EMIC is about 10% larger than that of the Schwarzschild's anistropic medium.

The results of ray-tracing are in agreement with the well-known images simulated in the literature [24,25]. In summary, the EMIC due to (5) would significantly over-estimate the true deflection of light in SAEM as given by (15). Finally, the simulated photographs of gravitational lensing due to the SAEM is presented in Fig. 5, obtained by digital graphical post-processing on Fig. 4.

## IV. OPTICAL INTERFEROMETRY

A simple plane gravitational wave propagating along $x$ axis with the angular frequency $\Omega$, wavenumber $K$, and amplitude $\delta$ is [12,26]

$$ds^2 = -c^2 dt^2 + dx^2 + \qquad (35)$$
$$\left[1 + \delta \sin(Kx - \Omega t)\right] dy^2 +$$
$$\left[1 - \delta \sin(Kx - \Omega t)\right] dz^2$$

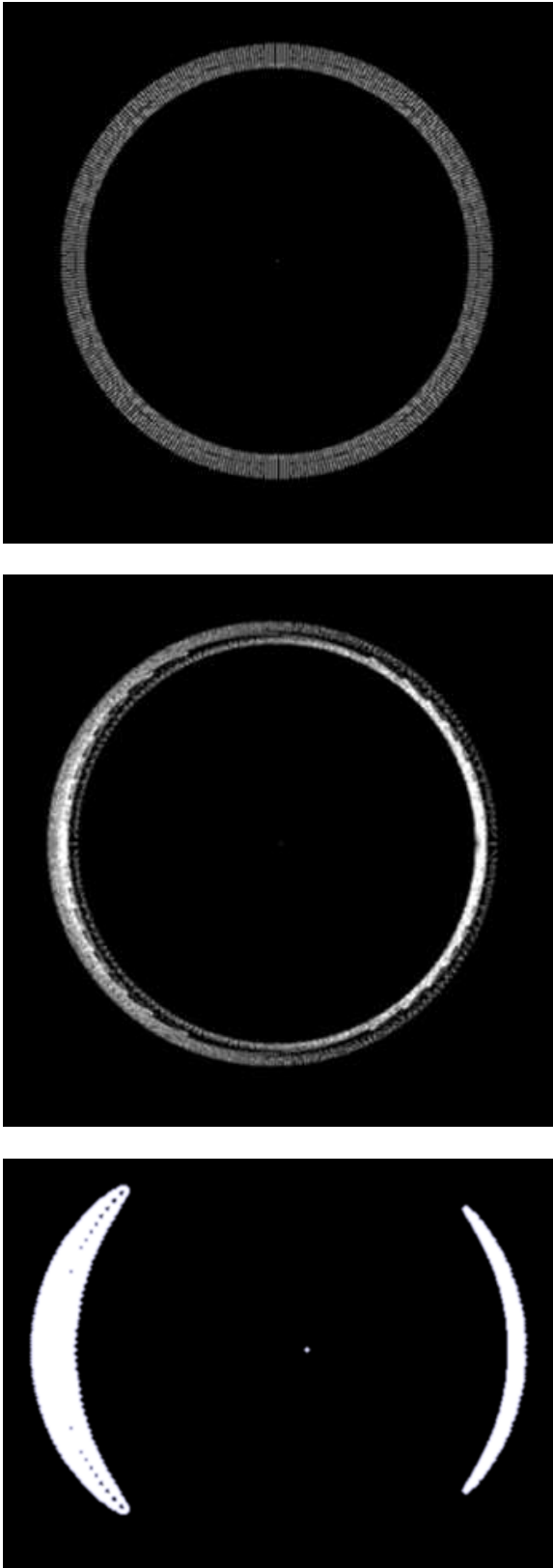

**Figure 5**. Simulated photographs of gravitational lensing due to the SAEM of Fig. 4.

The metric (35) is obviously anisotropic, and following [11,12] we obtain a maximum anisotropy of

$$\Delta n = n_y - n_z \approx \delta \sin(\Omega t) \qquad (36)$$

which is expected to be of the order of $10^{-24}$ [26]; this is yet to be observed in LIGO [26] and the corresponding recent upgrade [27].

As it can be seen, the fundamental operation of LIGO relies on the temporal geometrical anisotropy due to the passing gravitational field. Since Schwarzschild metric also causes such anisotropy, interferometry should be equally applicable to probe static gravitational fields. As stated above, the expected order of anisotropy caused by Sun is quite appreciable, being about $10^{-8}$, which is at least 16 orders of magnitude stronger than that of typical gravitational waves. This figure is large enough to be easily detectable on a tabletop setup. If it would, then most normal optical setups had problems working on the Earth, due to interference with the gravity of Sun, Earth, and other massive objects nearby. This is clearly not the case, and as the matter of fact, nobody has noticed such a large deviation in interferometric experiments.

Classically, all experiments in search of optical anisotropy of space using optical interferometry (and other methods) have failed [28-36] to reveal the existence of any appreciable anisotropy. The proposal of OPTIS satellite [37] suggests a three-orders-of-magnitude improvement over earlier interferometric Michaelson-Morley experiments. What is notable in all the reported interferometric experiments is that they all used rotating optical tables, which had their planes of rotation in parallel to the surface of the Earth. This could result in observation of no change in refractive index as predicted by (7), and hence no gravity-induced anisotropy. For instance, in order to observe a change in refractive index due to gravity of the Earth, it would be necessary for the interferometer arms to rotate in a plane normal to the surface of the Earth. An interferometric experiment done as such, would rigorously establish the possibility of any gravity-induced optical anisotropy.

On the other hand, if interferometry fails to reveal the existence of Schwarzschild metric (which is the apparent case), then one could expect the gravitational waves not to be detectable, too. This might explain why the LIGO experiment, despite its extremely high accuracy, has been unsuccessful in the search of gravitational waves.

**A.** *Fiber Optical Gyroscope Interferometer*
Here, we discuss an idea for implementation of a local interferometer based on the combination of two laser Fiber Optical Gyroscopes (FOGs). An FOG introduces a measurable phase shift between two counter-rotating beams. A relativistic calculation then gives the expression for the phase shift of a standard FOG as [38]

$$\Theta = \frac{8\pi NS\Lambda P}{\lambda c\sqrt{1-\left(R\Lambda/c\right)^2}} = \frac{4\omega NS\Lambda P}{c^2\sqrt{1-\left(R\Lambda/c\right)^2}} \quad (37)$$

where $\Theta$ is the phase shift, $N$ is the number of turns, $\Lambda$ is the mechanical angular frequency of rotation, $P$ is a constant, $S$ is the cross-sectional area of the fiber winding, $R$ is the radius of the winding cylinder, and $\lambda$ is the wavelength of the laser having the angular frequency $\omega$. Furthermore, $c$ is the local speed of light normal to the axis of mechanical rotation, i.e. on the propagation plane.

Now, we consider two identical FOGs, each having only one winding. These two FOGs may be mechanically coupled, so that they have the same angular frequencies $\Lambda_1 = \Lambda_2$. Suppose that the axes of rotation for these two FOGs are perpendicular, one resting on the plane normal to the local gravitational field. Then, if any anisotropy having the form (7) holds, one could expect that $c_1 \neq c_2$. We evidently have $c\left|c_1^{-1} - c_2^{-1}\right| = \left|n_1 - n_2\right| \approx \tfrac{1}{2}\left|\Phi\right|$. Hence, if $R\Lambda \ll c$ holds, the phase shift per each turn of fiber winding between the two rotating fiber windings would be given by

$$\frac{\Delta\Theta}{N} \approx 4\omega S\Lambda P\left|c_1^{-2} - c_1^{-2}\right| = \frac{4\omega S\Lambda P}{c^2}\left|\Phi\right| \quad (38)$$

which shows that a measurable phase shift in case of any possible local anisotropy must develop proportional to the product of mechanical rotation frequency $\Lambda$ and local gravitational potential $\left|\Phi\right|$. It is not difficult to anticipate the order of this phase shift: taking $P$ of the order of unity, $S$ of the order of 10cm$^2$, $\omega$ of the order of $10^{16}$rad/sec, and $\Lambda$ of the order of $10^4$rpm, we get a phase shift $\Delta\Theta/N$ of the order of $\left|\Phi\right|/N$. Now, typical value of $\left|\Phi\right|/N$ at the surface of earth is ranging from $10^{-10}$ to $10^{-8}$, which results in a typical phase shift range of the same order for $\Delta\Theta/N$. Increasing the angular mechanical velocity, number of turns, or a choice of shorter wavelength might help to increase the sensitivity of the experiment.

## V. CONCLUSIONS

Integration of ray-tracing equations obtained using transformation optics technique, has enabled us to compare the images of gravitational lensing in the isotropic and Schwarzschild's anisotropic equivalent media. It was established that anisotropy of Schwarzschild metric causes a first-order correction in terms of the gravitational potential, which causes a significant reduction in the apparent size of the image. This would mean that Einstein's expression for light deflection could significantly over-estimate the blackhole's mass. A discussion on the operation of optical interferometers in search of gravitational waves was presented, and it was pointed out that such experiments are all expected to detect the temporal anisotropy of spacetime caused by passing gravitational waves. Such anisotropy due to the gravities of the Sun and the Earth might already exist at the surface of the Earth, stronger by respectively sixteen and fourteen orders of magnitude. Hence, it was argued that optical interferometry might have been the inappropriate choice for detection of gravitational waves. Otherwise, a local interferometer based on the combination of two laser fiber gyroscopes could reveal the existence of any such anisotropy in the gravitational field.

**Farhad Karimi** was born in Tehran, Iran on November 2, 1987. He received his B.Sc. degree in Electrical Engineering from Sharif University of Technology in September 2010, where he is an M.Sc. student of Electrical Engineering since then. His research interests include quantum optics and transformation optics.

**Sina Khorasani** was born in Tehran, Iran, on November 25, 1975. He received the B.Sc. degree in electrical engineering from Abadan Institute of Technology in 1995, and the M.Sc. and Ph.D. degrees in electrical engineering from Sharif University of Technology, in 1996 and 2001, respectively. After spending a two-year term as a Postdoctoral Fellow with the School of Electrical and Computer Engineering, Georgia Institute of Technology, Atlanta, he returned to Sharif University of Technology, where he is currently an Associate Professor of electrical engineering in the School of Electrical Engineering. His active research areas include photonics, quantum optics and electronics. He is a Senior Member of IEEE.